\magnification=1200 \tolerance=5000
\hsize = 16.0 truecm \vsize = 23.6 truecm
\def\thf{\baselineskip=\normalbaselineskip\multiply\baselineskip
by 4\divide\baselineskip by 3}
\thf
\pageno=0

\def\nabl{\nabla\!}
\def\vex{\overrightarrow}
\def\el{\ell}
\def\spose#1{\hbox to 0pt{#1\hss}}
\def\Libra{\spose {--} {\cal L} }
\def\Diam{\spose {\raise 0.3pt\hbox{+}} {\diamondsuit}  }
\def\diam{\diamondsuit}
\def\diaS{\diam_{\rm S}}
\def\diaL{\diam_{\rm L}}

\def\ov{\widetilde}

\def\omic{\perp}

\def\DD{\delta}
\def\DL{\delta_{\rm L}}
\def\DE{\delta_{\rm E}}
\def\DS{\delta_{\rm S}}
\def\nnu{\nu}
\def\eth{\eta}

\centerline{\bf COLD, WARM, AND COMPOSITE (COOL)}
\centerline{\bf COSMIC STRING MODELS}

\vskip 1.0 cm
%\impd
\centerline{Brandon \bf Carter}

\vskip 0.6 cm
%\impb
\centerline{ D\'epartement d'Astrophysique Relativiste
et de Cosmologie, C.N.R.S.,}
\centerline{Observatoire de Paris, 92 195 Meudon, France.}
\bigskip\vskip 0.2 cm
%\impr
\centerline{(1993 preprint, submitted to {\it Nucl. Phys.} {\bf B}.)}
\vskip 1.0 cm
 \parindent = 0 cm

{\bf Abstract.} The dynamical behaviour of a cosmic string is strongly affected
by any reduction of the effective string tension $T$ below the constant value
$T=m^2$ say that characterizes the simple, longitudinally Lorentz invariant,
Goto Nambu string model in terms of a fixed mass scale $m$  whose magnitude
depends on that of the Higgs  field responsible for the existence of the string.
Such a reduction occurs in the standard ``hot" cosmic string model in which the
effect of thermal perturbations of a simple Goto Nambu model is expressed by the
formula ${T^2}=$ $m^2(m^2-2\pi\Theta^2/3)$, where $\Theta$ is the string
temperature. A qualitatively similar though analytically more complicated
tension reduction phenomenon occurs in ``cold" conducting cosmic string models
where the role of the temperature is played by an effective chemical potential
$\mu$ that is constructed as the magnitude of the phase $\varphi$ of a bosonic
condensate of the kind whose existence was first proposed by Witten.  The
present article describes the construction and essential mechanical properties
of a category of ``warm" cosmic string models that are intermediate between
these ``hot" and ``cold" extremes. These ``warm"  models are the string
analogues of the standard Landau model for a 2-constituent finite temperature
superfluid, and as such involve two independent currents interpretable as  that
of the entropy on one hand and that of the bosonic condensate on the other. It
is surmised that the stationary (in particular ring) equilibrium states of such
``warm" cosmic strings may be of cosmological significance.

\vfill\eject

\bigskip \parindent = 0 cm
{\bf1. Introduction: the importance of longitudinal Lorentz symmetry breaking. }
\medskip\parindent= 2 cm

As a model for the representation of cosmic string behaviour, i.e. for the
description at a macroscopic level of the phenomen whose underlying mechanism is
that of a vortex defect of the vacuum (in a field theory with spontaneous
symmetry breaking), the simplest possibility is specifiable by an variation
principle of the well known Goto-Nambu type, which means that the internal state
of its 2-dimensional world sheet is locally Lorentz invariant:
with respect to an internal rest frame whose choice is (as an expression of the
invariance property) entirely arbitrary the mass-energy density per unit length
$U$ and the associated string tension $T$ will actually be equal, being given in
units with $\hbar=c=1$, simply by
$$ U=T=m^2 \, \eqno(1.1)$$
while the dynamical motion of such a string will be governed by an action given
as the integral over the string world sheet of a scalar Lagrangian function $L$
of the trivially simple form
$$ L=-m^2 \ , \eqno(1.2)$$ 
where $m$ is a fixed mass scale which (as was pointed out by Kibble in the
earliest discussion of the potential cosmological importance of the
phenomenon$^{[1]}$) may be expected to be of the same order of magnitude as the
mass scale characterising the Higg's boson responsible for the spontaneous
symmetry breaking, the precise value (as derived by working out the
corresponding Nielsen Olesen$^{[2] }$ type equilibrium state) being dependent on
the details of the particular underlying field theory that is supposed to be
relevant. In most of the early discussions of the subject it was postulated that
the strings under consideration were of the ``heavyweight" variety for which the
relevant symmetry breaking was that of grand unification, as characterised by
$Gm^2\approx 10^{-6}$, but (as has recently been shown explicitly by Peter$^{[3]
}$) it is also possible to envisage ``lightweight" cosmic string formation for
which the relevant symmetry breaking is that of electroweak unification, as
characterised by $Gm^2\approx 10^{-32}$.

A very important consequence of the longitudinal Lorentz symmetry condition
expressed by (1.1) is that an isolated string loop of such a Goto-Nambu type has
{\it no stationary equilibrium} state and so must oscillate until all its energy
has been lost by gravitational radiation. However, for a generic string model,
in which this special local symmetry will be broken, such an energy loss process can
not go through to completion due, as was first clearly pointed out by by Davis
and Shellard$^{[4][5][6]}$, to the consequent existence of stationary
equilibrium states. The recognition of this effect immediately leads to the
question -- which does not arise in the pure Goto-Nambu case -- of whether a
cosmological mass density excess ${\mit\Omega}\gg 1$ would not have been produced
by the formation of a distribution of stationary loop relics, at
least$^{[4][5][6]}$ for strings of the  ``heavyweight" variety characterised by
$Gm^2\approx 10^{-6}$, which makes it possible to argue$^{[7][8]}$ that strings
of the ``lightweight" variety characterised by $Gm^2\approx 10^{-32}$ are more
likely (or less unlikely) than the ``heavyweight" variety to have played a real
role in the evolution of our universe. 

It is to be remarked  that even before the key point was made by Davis and
Shellard$^{[4][5][6]}$ it had already been pointed out that string loops
might be maintained in equilibrium by a global as opposed to local
mechanism, namely that of magnetic support$^{[9]}$, in which the tendency to
contraction due to the local string tension $T$ is balanced by the the
effect of a globally extended dipolar magnetic field due to an
electromagnetically coupled current in the loop. It was however recognised
at the outset$^{[9]}$ and agreed in subsequent discussions$^{[10][11][12]}$
that (taking account of the smallness of the electromagnetic coupling
constant) such an electromagnetic support mechanism was unlikely to be
sufficiently strong to be of practical importance. The subsequent
discussions gave rise nevertheless to a certain amount of confusion due to
their failure to clearly distinguish this essentially {\it global} magnetic
support mechanism (due to the long range effect of the external
electromagnetic field arising from the charge coupled current) from the
qualitatively different mechanism (which can also be considered to account
for the phenomenon of current saturation, and which operates even in the
absence of any electromagnetic coupling) whereby the {\it local} string
tension is reduced by the mechanical effect of the current in the immediate
neighbourhood of the vortex core. This confusion was embodied in the use of
the potentially misleading term ``spring" to describe states in which the
effective tension was reduced to zero so as to allow static equilibrium
without an external supporting force. More detailed examination$^{[13][14]}$
of the local effect of the current appeared to confirm that in certain
models the string tension $T$ actually could in principle reach zero and
even negative values, but it was pointed out that in a local (as opposed to
global) string state negative tension automatically implies
instability$^{[15]}$ and it has been shown more recently using improved
numerical methods$^{[16][17]}$ that although it can be considerably reduced,
the local string tension $T$ will in fact remain strictly positive in the
all the kinds of model that have been examined so far. 

However that may be, the question of whether purely magnetically supported
equilibrium states can be of of any practical significance (and of whether local
``spring" states can exist at all) was effectively relegated to obsolescence as
far as cosmological implications are concerned by the observation of Davis and
Shellard$^{[4][5][6]}$ that there is a fundamentally different, essentially
local rather than global support mechanism that is is quite independent of
electromagnetic coupling and that will in any case be {\it much more effective},
namely the centrifugal effect that can operate whenever the local longitudinal
Lorentz symmetry (1.1) is violated so as give a strict inequality,
$$  T<U \ , \eqno(1.4)$$
since this is all that is required to allow existence of  centrifugally
supported equilibrium states (of which the simplest$^{[18][19]}$, though not the
only type$^{[20][21]}$, are circular  ring configurations) which are
characterised by a subluminal longitudinal running speed v<1  that is given
quite generally by
$$v^2={T\over U} \ .\eqno(1.5)$$

\bigskip\parindent=0cm
{\bf  2. Lorentz symmetry breaking by Witten's ``cold" current mechanism.}

\medskip\parindent=2 cm

The development that originally raised the question of the cosmological
implications of longitudinal Lorentz symmetry breaking and the consequent
existence of  centrifugally supported equilibrium states
 was the demonstration by Witten$^{[22]}$ that for field theories
of a kind only slightly more complicated than that on which Kibble's original
analysis was based, the core of the vortex defect of the vacuum may be inhabited
by a boson condensate of the type familiar in ordinary superfluid theory with
the implication that it may support a corresponding dissipationless current. At
a macroscopic level the resulting ``superconducting" (or to be technically more
precise ``charged superfluid") string behaviour  will be describable by a type
of model that is specifiable$^{[18][23]}$ by replacing the degenerate (constant)
Lagrangian (1.2) by a function of a {\it variable} mass, $\mu \approx m$,
obtained as the magnitude of an effective energy-momentum covector $\mu_\rho$
that is itself constructed from the gradient of a scalar phase field $\varphi$.

In cases for which no coupling to any external field need be taken into account,
the energy-momentum covevctor will be given simply by
$$ \mu_\rho=\ov\nabl_{\!\rho}\varphi\ ,\eqno(2.1)$$
where the $\ov\nabl_\mu$ denotes the worldsheet projected gradient 
operator$^{[18][23]}$ as given by
$$ \ov\nabla_{\!\rho} = \eta_\rho{^\sigma}\nabla_{\!\sigma} \eqno(2.2)$$
in terms of the fundamental$^{[18][23]}$ tensor
$\eta_\rho{^\sigma}$
(whose definition is given later on in section 5). 

The form of Lagrangian function
$$ L=L(\mu)\ , \hskip 2cm \mu^2=-\mu_\rho \mu^\rho\ \eqno(2.3)$$
that is to be used in place of (1.2) will remain unaffected by allowance for
charge coupling to an external electromagnetic field, such as was postulated
in Witten's original discussion$^{[22]}$, the only effect of such a
coupling, when relevant, being to modify the defining formula for the energy
momentum covector $\mu_\rho$ by the requirement that the appropriate gauge
covariant differentiation operator $D_\rho$ should be used instead of
$\nabla_\rho$. Such secondary external coupling effects will not however be
considered in the present discussion, partly because they would divert
attention from the new concepts that will be presented, and partly because
they no longer seem quite to be of quite such dominant importance as was
once supposed.

Actually, although great importance was attributed to charge coupling
effects in the earliest discussions of the Witten mechanism$^{[9]}$  it has
become apparent since the work of Davis and Shellard$^{[4]}$, and
particularly since the recent work of Peter$^{[17]}$ that they are 
relatively less important than the purely mechanical effects to which our
attention will be restricted in the present work: the smallness of the
electromagnetic ``fine structure" coupling constant $e^2=1/137$ justifies
its neglect, at least as a lowest order approximation, in a wide range of
circumstances, which in the context of conducting strings has the
significant technical advantage of ensuring that there is no need to worry
about the subtle infrared divergence problems (violation of the locality
postulate on which a string description in the strict sense is based) that
would otherwise arise. Subject to this condition, i.e. assuming either the
actual  absence or the effective negligibility of any electromagnetic
coupling, Peter has carried out accurate numerical calculations (superseding
earlier more approximate investigations$^{[14]}$) of the equation of state
specifying the non-linear Lagrangian to be used in (2.3) on the basis of a
simplified toy bosonic field theory$^{[16]}$ (with two complex scalar fields
but just one gauge boson) whose complexity is midway between the one
originally used by Kibble$^{[1]}$ for the non conducting case (with just one
complex scalar field and one vector gauge boson) and the one originally
proposed by Witten$^{[22]}$ (which had not only two complex scalar fields
but also two independent vector gauge bosons). 

Although Witten originally raised the question of conductivity in cosmic strings
on the basis of  a model that was too highly simplified to be realistic, a large
number of other conceivable (fermionic as well as bosonic) ``superconduction"
mechanisms have since been suggested by many authors in less artificially
idealised models (of which one of the most realistic is perhaps the kind
recently discussed by Peter$^{[3]}$ within the framework of the modified
electroweak unification theory that has been proposed by Fayet$^{[24]}$.)
The existence of such a rich range of possibilities suggests that far from
being exceptional, ``superconductivity" of one kind or another is likely to
be a generic feature of cosmic strings if they exist at all, or to put it
another way, the postulate of its total absence would seem to be a severely
restrictive condition to impose on any would-be realistic cosmic string
forming field theory.

Unless (conceivably as an application of a sufficiently strong version of the
anthropic principle$^{[25]}$) a specially contrived  non superconducting string
forming theory  actually does apply, the expected currents will determine
corresponding conserved quantities$^{[18][19][23]}$ on any closed string loops
that may be formed, which -- provided the relevant length scales are large
enough for quantum tunnelling to be negligible so that a purely classical
description remains valid -- will exclude the possibility of their complete
destruction by radiative energy loss so that the stationary relic formation
process first envisaged by Davis and Shellard$^{[4]}$ will inevitably occur. As
was remarked in the preceeding section, for ``heavyweight" (G.U.T. mass scale)
strings as characterised by $Gm^2\approx 10^{-6}$, simple dimensional
considerations$^{[4][5][6][7][8]}$ lead to the prediction of such an extremely
large cosmological mass density excess ${\mit\Omega}\gg 1$  that even if the
efficiency of the relic loop formation were very low it would still be hard to
obtain a realistic scenario with ${\mit\Omega}\approx 1$ unless the classical
description is invalidated at some stage. As a way of saving the scenario of
galaxy formation by heavyweight cosmic string formation, Davis and Shellard
suggested$^{[4][5]}$ that it might after all be possible to get rid of nearly
all the  unwanted relic loops by quantum tunnelling to non-conducting Kibble
type states. However although such tunnelling processes will no doubt have some
effect, it is evident that the corresponding decay timescales  will be very
sensitively dependent on the length scales and wavelengths that are involved,
and hence indirectly on the values of the classically conserved numbers by which
the loops are characterised. The random processes by which the loops are formed
will presumably give rise to widely scattered values for the corresponding
conserved numbers, so that, although many or even most may fall in a range for
which the corresponding tunnelling timescales are cosmologically short, the
existence of a systematic mechanism to prevent even a small fraction from being
stable over cosmologically long timescales is not at all obvious. The
implication is that the viability of the heavyweight string scenario is much
more seriously threatenned by the loop preservation mechanism of Davis and
Shellard  than these authors themselves originally realised. 

On the other hand, although some rather miraculous (e.g. anthopic) deus ex
machina process would seem to be needed to restore the credibility of string
formation scenarios with $Gm^2\approx 10^{-6}$,  it is nevertheless to be
emphasized that there is no such difficulty for lightweight string scenarios as
characterised by the electroweak value $Gm^2\approx 10^{-32}$ which appears on
the basis of plausible assumptions$^{[7][8]}$ to be just high enough to allow
the ensuing distribution of relic loops  to provide a cosmologically significant
mass distribution, ${\mit\Omega}\approx 1$, in the form of what would presumably
be perceived as a cold dark matter contribution.
\vfill\eject

\bigskip\parindent=0 cm
{\bf 3. The standard ``hot" string model.}
\medskip\parindent =2 cm

The main purpose of the present work is to draw attention to, and provide a
mathematical framework for analysing, another effect of a different nature to
those discussed in the previous sections, which also effectively contributes to
longitudinal Lorentz symmetry breaking and therefore enhances the efficiency of
the loop preservation phenomenon whose potential cosmological importance has
been underlined above. The effect in question is that of stochastic, and more
particularly {\it thermal} excitation not just of internal excitation modes
within the string, whose energy thresholds may conceivably be too high for them
to be important, but more particularly of the {\it extrinsic perturbation} modes
whose potential relevance can not be doubted, since their existence is already
implicit in any string description at all, including  the ultra simple
Goto-Nambu model characterised by (1.2) as well as the less degenerate models
characterised by (2.3).

Even in the case of a model of the simple non-superconducting type originally
proposed by Kibble$^{[1]}$, a more accurate treatment would require allowance
for the excitation of shortwavelength modes which would cause deviations from
the simplified description based just on the constant Goto Nambu action (1.2).
If the spectrum is thermal and if the only modes that are excited are extrinsic
transverse displacements of the string worldsheet itself then a simple heuristic
argument lead me to the prediction$^{[26]}$ that the motion will still be
describable by a variation principle, but one in which the constant (1.2) is
replaced by an effective Lagrangian that is a function of the relevant surface
density current of entropy, $s^\rho$ say, whose existence evidently breaks the
longitudinal Lorentz symmetry by specifying a corresponding preferred thermal
rest frame. When expressed in energy units (such that Boltzmann's constant is
set to unity, i.e. k=1) the explicit analytic form of the predicted Lagrangian is
given by the simple expression
$$L = -m^2\Big\{1+{3  s^2\over 2\pi  m^2}\Big\}^{1/2}\ ,
\hskip 2cm     s^2=-s^\rho s_\rho \ . \eqno(3.1)$$
where the scalar $s$ is interpretable as the surface density of entropy in the
thermal rest frame. In the application of the variation principle the surface
current vector $s^\rho$ is to be varied not quite freely but,
 subject (as discussed in detail in the Section 5) to a
constraint ensuring that its surface divergence, $\ov\nabla_\rho s^\rho$ 
vanishes. 

The resulting mechanical behaviour in this standard ``hot" cosmic string
model is describable in terms of a thermal rest frame surface energy $U$ and
a corresponding string tension $T$ that are given by
$$U={m^4\over T}=-L \ . \eqno(3.2)$$
The qualitative conclusion that the product $UT$ remains constant is not
dependent on the supposition that the excitation ``noise" spectrum is
strictly thermal, and has been confirmed by a more detailed and explicit
calculation of the effect of small ``wiggles" by Vilenkin$^{[27]}$. The
``constant product" equation of state (3.2)  can be shown to imply direct
{\it integrability} of the dynamic equations not only in a flat
background$^{[26]}$ but also for the case of stationary equilibrium in a
Kerr black hole background$^{[19]}$, the underlying reason for all these
exceptionally convenient properties being that this particular kind of
equation of state is uniquely characterise  by equality of the extrinsic and
longitudinal characteristic speeds $c_{\rm _E}$ and $c_{\rm _L}$ as
given$^{[15]}$ by
$$  c^2_{\rm _E}={T\over U}\ , \hskip 2 cm  c^2_{\rm _L}=-{dT\over dU}
\eqno(3.4)$$
It is to be remarked that behaviour characterised by the same ``constant
product" equation of state (3.2) can also be shown$^{[18]}$ to arise in the
physically very different and rather artificial context of Nielsen's Kaluza
Klein mechanism$^{[28]}$.

In the natural thermal case with which we are concerned here, the corresponding
local {\it temperature} of the string, which it is convenient to denote by
$\Theta$ (to avoid confusion with the tension $T$) will be obtainable from (3.1)
by application of the standard defining relation 
$$\Theta={dU\over ds} \eqno (3.5)$$ 
which leads to the relation 
$${2\pi\Theta^2\over 3 m^2}=1-{T^2\over m^4}  \ .\eqno(3.6)$$ 
It can be seen that starting from the longitudinally Lorentz invariant (Kibble
type) state characterised, in accordance with the Goto-Nambu limit formula
(1.1), by $U=T=m^2$, the surface energy density $U$ increases without limit as
the tension $T$ decreases towards zero, while the temperature $\Theta$ also
increases monotonically as $T$ decreases, though not without limit but only
towards a finite maximum (Hagedorn type) saturation value $\Theta_{\rm max}$
that is of the same order of magnitude as the relevant cosmological symmetry
breaking phase transition temperature at which the string formation occurred in
the first place: it can be seen from (3.6) that explicitly, in terms of the
relevant (Higgs) mass scale $m$, this maximum saturation temperature will be
given by 
$$\Theta_{\rm max}=\sqrt{3\over2\pi}\, m\ , \eqno(3.7)$$

It is to be expected that the cosmic strings should originally have been formed
with a temperature of effectively the same order as the Higgs mass scale
characterising the symmetry breaking phase transition involved, i.e. with
$\Theta\approx\Theta_{\rm max}$, but that they would at first have been strongly
coupled to their environment and would therefore have cooled rapidly as the
universe expanded so that by the time they emerged as effectively free systems
governed without allowance for external drag corrections just by the Goto-Nambu
action (1.2) in so far as small scale behaviour is concerned and by the
standard ``hot string" Lagrangian (3.1) at the level of larger scale
averaged behaviour, the temperature would have been brought down to a very
much lower value, $\Theta\ll \Theta_{\rm max}$. However although this
implies that the in the short run the mechanical effects of the thermal
current would be relatively small, it is by no means obvious that they would
not become important in the long run provided the heat loss by thermal
radiation is sufficiently small for the entropy current to be sufficiently
well conserved over timescales long enough for a loop to have undergone
significant overall contraction due to long wavelength gravitational
radiation, since under such circumstances the current would ultimately
become more densely concentrated with the consequence that the string would
tend to heat up again.  A quantitative analysis of the likely efficiency of
this potentially important reheating effect is therefore an interesting
subject for future work.

\vfill\eject 

\bigskip\parindent=0 cm
{\bf 4. The need for composite ``warm" cosmic string models.  }
\medskip\parindent=2 cm

It is apparent that there is a sense in which the ``hot" string model
characterised by (3.1) -- in which the only current is of entirely thermal
origin -- represents an opposite extreme from the ``cold" limit represented by
the Witten type ``superconducting" (or in stricter terminology ``superfluid")
models characterised by a Lagrangian of the form (2.3) -- in which there is no
allowance for any thermal contribution at all. Models of the latter type are
interpretable as the 1-dimensional analogues of an ordinary 3-dimensional
superfluid of the simple single constituent type that is exemplified by
ordinary liquid helium-4 in the limit of {\it strictly zero} temperature. It is
however well known that when one needs to take account of the effects of
{\it finite} temperature in liquid helium-4 one has to use a model of the two
constituent type whose prototype was first set up by Landau. The essential
feature of Landau's ``warm" superfluid model$^{[29]}$ is the coexistence of
two independent (but of course strongly coupled) currents representing the
independent degrees of freedom of a ``cold" constituent, corresponding to
the boson condensate, on one hand, and of a ``hot" (or in the traditional
technical jargon ``normal") constituent, corresponding to the entropy
current (which at low temperatures can be accounted for mainly in terms of
``phonons"). In the context of cosmic strings, the analogue of the ``cold"
part is the Witten type condensate discussed in Section 2, while the
analogue of the ``hot" part is the thermal excitation current discussed in
section 3. Our purpose here is to describe the way to allow for the presence
of both kinds of current together in a ``warm" superfluid cosmic string
model that will be the 1 dimensional analogue of the ordinary 3-dimensional
Landau model. 

Before proceeding, it is to be remarked that
the simple explicit formulae presented in the previous section allow for the
macroscopic thermal description only for what might be described as ``snake"
modes but not for other more complicated possibilities (of which the most
obvious are what might be described ``sausage" modes) whose inclusion in a more
accurate warm string model would presumably require numerical not just analytic
calculations. When we start not from an underlying smallscale string that is not
of the Goto Nambu type characterised by (1.2) that was considered in section 3
but of the ``superconducting" type characterised by (2.3), the situation becomes
even more complicated, because an explicit analytic treatment is not available
even for the unperturbed equation of state, and hence a  fortiori not for the
thermal treatment of even the simplest ``snake" type perturbation modes.
Such a state of relative ignorance need not however discourage us from
developing a formal framework for the qualitative analysis of variables whose
explicit quantitative values one may hope to obtain later from detailed
numerical computations, since this state of affairs is closely analogous to that
which pertains to the familiar laboratory phenomenon of superfluidity in liquid
helium - the only essential difference being that in the laboratory case results
that are beyond the scope of present day numerical analysis may already be
empirically accessible to some extent by experimental methods. 

In the laboratory case a considerable body of experience encourages faith in the
use, as a first approximation, of the standard Landau theory $^{[29]}$ whose
natural relativistically covariant generalisation of this theory has recently
been shown$^{[30]}$ to have a variational formulation in terms of a Lagrangian
of the form
$$ L = L (\mu_\rho, s^\rho) \ , \eqno(4.1)$$
that is specified as a scalar function of
field variables that are taken to be a gradient $\mu_\rho$
constructed, as in (2.1), from the scalar phase
$\varphi$ of an underlying quantum condensate (as in the single current 
string model governed by (2.3)) and an entropy current vector $s^\rho$
(as in the hot cosmic string model governed by (3.1)) whose variation is
not considered to be free but subject to a constraint of convective 
type$^{[31]}$. Of course if very high accuracy were required such a strictly
conservative treatment would not be quite sufficient and would need to be
modified to allow for effects such as viscosity of the entropy flux, but its
sufficiency for many purposes, including the description of first and second
sound perturbation modes, encourages its translation to the highly analogous
(theoretically rather simpler though experimentally less accessible) context of
cosmic strings for which the zero temperature limit, $s^\rho=0$ and the non
``superconducting" limit $\mu_\rho=0$ are already familiar as the special cases
given respectively by (2.3) and (3.1).  

The way to do this is described in the following sections, in which it will be
shown how to set up a class of ``warm superfluid" $p$-brane models that includes
the relativistic version of the ordinary Landau two constituent superfluid model
in the case $p=3$ and that includes the ``warm" conducting string model in which
we are particularly interested as the case $p=1$. Since this model itself
includes the original Witten type string model as a zero temperature ``cold"
limit, it can be seen that the fashionable use of the term ``superconducting" is
rather an exageration for describing this particular case since it corresponds
in this categorisation to the simplist kind of superfluid as exemplified by zero
temperature helium and not to a proper superconductor of the kind exemplified by
many low temperature metallic media. (Such genuinely superconducting behaviour
is indeed likely to be exhibited by cosmic strings of the more complicated type
that arise from more realistic theories such as that of Peter$^{[3]}$, but a
simple Witten type model of the form characterised by (2.3) has too few degrees
of freedom for superconductivity in the strict sense to be able to occur, and
would therfore be more appropriately described as a ``charged superfluid" string
so as to leave the term ``superconducting" string avilable for distinguishing
the more complicated cases that remain as an interesting subject for future
investigation).

\bigskip\parindent=0 cm
{\bf 5. Entropy flux in a conservative $p$-brane model.}

\medskip\parindent=2 cm

In order to set up a category of models that manifestly includes both the
relativistic version of the standard Landau two-constituent superfluid theory
and also  the warm ``superconducting" cosmic string model in which we are most
particularly interested as a special cases, we shall start by considering a
$p$-brane of arbitrary dimension, using the standard convention$^{[32]}$ that
$p$ here refers to the dimension of a spacial section through the structure
under consideration, which means that the dimension of its spacetime world sheet
will be $p\!+\!1$. Thus the case of an ordinary superfluid will be specified by
$p=3$, and the string case in which we are primarily interested here will be
specified by $p=1$, while the intermediate case $p=2$ corresponds, at least in a
4 dimensional spacetime background, to the hypersurface supported case of a
membrane. For the purpose of the present section the background spacetime
dimension $n$ will not be restricted, so that hypothetical higher dimensional
cases are also implicitly included. The position of the brane worldsheet
relative to to local background spacetime coordinates $x^\mu\ ,$ 
$(\mu=0,1, ...n-1)$ (where in the usual applications $n=4$) will therefore 
be given by a mapping of the form                
$$\sigma^i\mapsto x^\mu(\sigma)\ .\eqno(5.1)$$ 
where the $\sigma^i$,  $i=0,1, ..., p$, are internal coordinates on the
$p$-brane world sheet. With respect to such a reference system the scalar field
$\varphi$ (representing the superfluid phase angle) on which the models set up
below will depend is simply given by another such internal coordinate mapping,
$$\sigma^i\mapsto \varphi(\sigma) \ . \eqno(5.2)$$

The essential feature of a ``warm" as opposed to ``cold" material medium or more
general $p$ brane model is a current of {\it entropy} which will be supposed to
be conserved in the not dissipative case to which the present discussion will be
restricted (as it must be if a variational treatment is to be obtained).  Such a
current can be represented mathematically by a  $p$-dimensional manifold 
$\cal X$ say with a corresponding $p$-surface measure $dS$ and local coordinates
$X^{\rm A}$ say, $(A=1, ..., p)$, that is obtained as the image of a projection
expressible in a form analogous to (5.1) and (5.2) as
$$\sigma^i\mapsto X^{\rm A}(\sigma)\ ,\eqno(5.3)$$ 
With respect to the comoving entropy current coordinates $X^{\rm A}$ the entropy
measure will be have an expression of the standard form 
$$dS=(p!)^{-1}S_{\rm A_1...A_p}dX^{\rm A_1}...dX^{\rm A_p}\eqno(5.4)$$
where the $S_{\rm A_1...A_p}$ are antisymmetric tensorial compenents that depend
only on the comoving coordinates $X^{\rm A}$ (and that can easily be made
uniform and normalised so as to have values $\pm1$ or $0$ by suitably adjusting
these comoving coordinates) so that they will determine a corresponding 
$(\!p\!+\!1\!)$ surface current $p$-form that will have components 
$$S_{i_1...i_p}=S_{\rm A_1...A_p}X^{\rm A_1}_{\ ,i_1}...
X^{\rm A_p}_{\ ,i_p}\eqno(5.5)$$
where a comma is used to indicate partial derivation with repect to the
coordinates whose indices are indicated, and that will automatically be closed
in the sense of having vanishing exterior derivative, i.e. 
$$(p+1)S_{[i_1...i_p,i_0]}=0\eqno(5.6)$$ 
using square brackets to denote index antisymmetrisation.   Since the world
sheet will have its own geometric measure tensor, with components 
${\cal E}_{i_0 i_1... i_p}$ say,  that is determined (modulo a choice of sign 
expressing the orientation) in the usual way by the square root of the modulus
of the determinent $\vert \eth\vert$ of the internal metric $\eth_{ij}$ , where the
latter is induced via (5.1) by the background metric $g_{\mu\nu}$ 
according to the usual pullback formula 
$$\eth_{ij}=g_{\mu\nu}x^\mu_{\ ,i}x^\mu_{\ ,j}\ ,\eqno(5.7)$$ 
it follows, raising internal indices by contraction with the inverse $h^{ij}$ of
the internal metric tensor $h_{ij}$, that the entropy current $p$-form will have
as its dual a corresponding entropy current vector 
$$s^i=(p!)^{-1}{\cal E}^{i j_1... j_p}S_{j_1...j_p}\eqno(5.8)$$ 
that will automatically satisfy the corresponding surface conservation law 
$$(\Vert \eth\Vert^{1/2}s^i)_{,i}=0\ .\eqno(5.9)$$ 

The introduction of the fundamental tensor $\eta^{\mu\nu}$ as
constructed$^{[15][18]}$ from the inverse $h^{ij}$ of the internal metric 
 according to the defining specification
$$\eta^{\mu\nu}=\eth^{ij}x^\mu_{\ ,i}x^\nu_{\ ,j}\ ,\eqno(5.10)$$ 
enables us to complete the construction of the surface projected gradient
operator $\ov\nabl_\nu$ as defined by (2.2) and hence to rewrite the
conservation identity (5.9) in terms of the the the corresponding background
current vector 
$$s^\mu=s^i x^\mu_{\, ,i} \eqno(5.11)$$ 
in background tensorial notation as
$$\ov\nabl_\mu s^\mu=0 \ . \eqno(5.12)$$

\bigskip\parindent=0 cm
{\bf 6. Lagrangian  displacement analysis in a brane model.}
\medskip\parindent=2 cm

The reason for going to the trouble of formally deriving the conservation law
(5.9) as a consequence of the more fundamental relations (5.3), (5.4), and
(5.5), rather than simply postulating postulating (5.9), or its tensorial version
(5.12), directly as an axiom in its own right is that the action principle that
we need is based on the free variation of the mappings (5.1), (5.2) and (5.3).
This means that unlike the scalar field $\varphi$, the entropy current vector
$s^\mu$ will not have the status of a independent field variable, its place
being effectively taken by the comoving coordinate fields $X^{\rm A}$. However
since the latter are not physically well defined, being subject to a very high
degree of gauge dependence in so much as they are replaceable by arbitrary
functions of themselves, it is desirable to introduce machinery whereby they can
be eliminated as soon as possible in favour of more physically meaningful
quantities such as $s^\mu$ itself.  All we have to do to achieve this is to take
over into brane theory the device whose use in the more familiar context of
ordinary fluid theory is associated with the name of Lagrange, i.e. to work with
a reference system that is itself comoving with respect to the variations to be
considered. Thus we can arrange that the only change in $s^i$ is the part due to
the variation of the measure, which just leaves
$$ \DL s^i=- {_1\over^2}s^i \eth^{jk}\DL \eth_{jk}\eqno(6.1)$$ 
simply by choosing internal coordinates $\sigma^i$ for the perturbed
system with respect to which the variations of the entropy space
coordinates reduce to zero, i.e. for which the mapping (5.3) is
preserved so that
$$\DL X^{\rm A}(\sigma)=0\eqno(6.2)$$
where  $\DL$ denotes the corresponding Lagrangian differential
operator associated with the infinitesimal change under consideration
(whereas if (6.2) were not imposed, the right hand side of (6.1)
would have to be supplemented by a lot of terms involving not only
the variations $\DL X^{\rm A}$ themselves but also their derivatives
$\DL X^{\rm A}_{\ , i}$).

Since there are only $p$ independent comoving coordinates $X^{\rm A}$, while the
$p$-brane world sheet is itself $(p\!+\!1\!)$ dimensional, it will be generically
possible while we are about it, consistently with (6.2), to use the remaining
freedom of internal coordinate adjustment to pin down the Lagrangian reference
system completely by taking it to be comoving with the phase field as well, i.e.
arranging to preserve the mapping (5.2) so as to have
$$\DL\varphi(\sigma)=0 \ .\eqno(6.3)$$
as well.

This still leaves us free to make what, for $p<n\!-\!1$ is the most important
simplification of all, which is to take the background coordinates to be
comoving with the $p$-brane world sheet itself, i.e. to freeze not just the
entropy current mapping (5.3) as is done by (6.2) and the phase mapping
(5.2) as is done by (6.3) but also to freeze the basis brane imbedding
mapping (5.1) by arranging to have
$$\DL x^\mu(\sigma)=0 \ , \eqno(6.4)$$
with the corollary that the corresponding variation of the
$(p\!+\!1\!)$ surface measure density will be given by
$$\Vert\eth\Vert^{-{_1\over^2}}\DL \Vert\eth\Vert^{_1\over^2}={_1\over^2}
\eth^{ij}\DL h_{ij}={_1\over^2}\eta^{\mu\nu}\DL g_{\mu\nu} \eqno(6.5)$$
while that of the (rank $p\!+\!1$) tangential projection tensor $\eta^\mu_{\
\nu}$ will be given in terms of its (rank $n\!-\!p\!-\!1$) complement, the
surface orthogonal projection tensor $\omic\!^{\mu}_{\ \nu}$ by
$$\DL \eta^\rho_{\ \sigma}=\eta^{\rho\mu}\!\omic\!^{\!\nu}_{\ \sigma}\,
\DL g_{\mu\nu}\eqno(6.6)$$
where the  latter is got in its fully
contravariant version $\omic\!^{\!\mu\nu}$ directly from the
fundamental tensor $\eta^{\mu\nu}$ as defined by (5.10) by
the defining relation
$$\omic\!^{\mu\nu}=g^{\mu\nu}-\eta^{\mu\nu}\ . \eqno(6.7)$$

The use of such a fully Lagrangian system can be seen from (5.7) and (5.10) to
have the advantage of making (6.1) immediately translatable into background
tensorial form as
$$\DL s^\rho= - {_1\over^2}s^\rho \eta^{\mu\nu}\DL g_{\mu\nu}\ .\eqno(6.8)$$
while for the momentum-energy covector $\mu_\mu$ constructed according to (2.1)
as the gradient of the phase scalar $\varphi$ one obtains the corresponding
relation
$$\DL \mu_\rho=\mu^\nu\omic\!^{\!\mu}_{\ \rho}\, \DL g_{\mu\nu} \ . \eqno(6.9)$$

\bigskip\parindent=0 cm
{\bf 7. Semi Eulerian ``surface sliding" displacement analysis 
in a brane model.}
\medskip\parindent=2 cm

Valuable though it is, the simplification (6.4) is seldom achieved without some
cost in practice, since the choice of the background coordinates $x^\mu$ is
usually not quite so arbitrary as that of the comoving current coordinates
$X^{\rm A}$, particularly in cases when gravity is only involved weakly or not at
all. It is usually the case that some criterion, such as flatness when gravity
is absent, determines a corresponding ``fixed" reference system which it would
be traditional to refer to as ``Eulerian", with corresponding local coordinates
coordinates $z^\mu$ say that could be used to fix the choice of the $x^\mu$ in
the unperturbed configuration, but which would be related to them by a
non-trivial infinitesimal displacement mapping determined by a corresponding
displacement vector field $\DD^\mu$ in the form 
$$x^\mu\mapsto z^\mu(x)= x^\mu+\DD^\mu(x)\eqno(7.1)$$
There is however no difficulty in translating from the Eulerian differential
$\DE$ say -- measuring the change of any background field quantity with respect to
such a reference system -- to the corresponding Lagrangian differential as denoted
by $\DL$, the difference being given$^{[33][34]}$ simply by the Lie derivative
$\vec{\DD\Libra}$ with respect to the relative displacement vector field
$\DD^\mu$ determined by (7.1), i.e. one has
$$\DL=\DE+\vex{\DD\Libra} \ .\eqno(7.2)$$

However although an Eulerian reference system may be indispensible for the
description of background spacetime fields, such as the metric itself
(particularly in cases where the background spacetime is fixed to be flat so
that one gets the convenient simplification $\DE g_{\mu\nu}=0$), on the other
hand an Eulerian description is useless for fields (such as $\varphi$ and
$s^\mu$ in our present case) whose support is confined to a lower dimensional
brane worldsheet. What can however be usefully done when worldsheet
displacements are involved is to introduce an intermediate ``surface sliding"
reference system obtained by a displacement of the form $$x^\mu\mapsto y^\mu(x)=
x^\mu+\zeta^\mu(x)\eqno(7.3)$$ where $\zeta^\mu$ is restricted to be {\it
tangential} to the worldsheet, i.e.
$$\omic\!^{\!\mu}_{\ \nu}\zeta^\nu=0 \eqno(7.4)$$
so that it will be obtainable, by applying the brane imbedding mapping (5.1) to
a corresponding internal worldsheet vector with components $\zeta^i$,  in the
form
$$\zeta^\mu=\zeta^i x^\mu_{\ ,i} \ , \eqno(7.5)$$
which means that there will be a corresponding infinitesimal internal coordinate
transformation
$$\sigma^i\mapsto\sigma^i+\zeta^i \eqno(7.6)$$
whose application in conjunction with (7.3) will preserve the form of the basic
imbedding mapping (5.1), thereby providing us with a new semi-Lagrangian,
semi-Eulerian, reference system (one that is Lagrangian from the point of view
of the background coordinates $x^\mu$ but that is Eulerian from the point of
view of the internal coordinates $\sigma^i$) which will characterise a
corresponding ``surface sliding" differential, $\DS$ say, that will satisfy a
differential imbedding conservation law 
$$\DS x^\mu(\sigma)=0 \ , \eqno(7.7)$$
of the same form as its fully Lagrangian analogue (6.4), with corollaries
analogous to (6.5) and (6.6), i.e.
$$\Vert \eth\Vert^{-{_1\over^2}}\DS \Vert\eth\Vert^{_1\over ^2}
={_1\over^2}h^{ij}
\DS \eth_{ij}={_1\over^2}\eta^{\mu\nu}\DL g_{\mu\nu}\eqno(7.8)$$
and
$$\DS \eta^\rho_{\sigma}=\eta^{\rho\mu}\omic\!^{\!\nu}_{\ \sigma}\,
\DS g_{\mu\nu}\eqno(7.9)$$
but which, unlike its fully Lagrangian anologue, will not be subject to
purely internal conservation conditions analogous to (6.2) and (6.3).

For our present purposes, the way of specifying such a semi-Eulerian
refence system that is most convenient is to postulate
that (instead of obeying the analogues of (6.2) and (6.3))
it should be characterised by the requirement that the further
{\it non tangential} displacement
$$y^\mu\mapsto z^\mu=y^\mu+\xi^\mu \eqno(7.10)$$
say, with $\xi^\mu$ specified by a  decomposition of the
form
$$\DD^\mu=\zeta^\mu+\xi^\mu\eqno(7.11)$$
that must be superimposed on (7.3) in order to get back
to (7.1), should satisfy the orthogonality condition that is
the obvious complement of (7.4), i.e.
$$\eta^\mu_{\ \nu}\xi^\nu=0.\eqno(7.12) $$
Having specified the decomposition (7.11) in this way, we can
think of the translation (7.2) as the result of successively performing
two subsidiary translations, namely a worldsheet orthogonal
displacement whose effect is expressible by
$$\DS=\DE+\vex{\xi\Libra} \ ,\eqno(7.13)$$
and a worldsheet tangential displacement whose effect is expressible by
$$\DL=\DS+\vex{\zeta\Libra} \ , \eqno(7.14)$$
of which the former is applicable only to background spacetime
fields such as the metric $g_{\mu\nu}$ itself,
 whereas the latter, (7.14) is equally well applicable also to fields
(such as $\varphi$ and $s^\mu$) whose support is confined to the brane
worldsheet.

The upshot of all this work is that the elimination of the variations of the
$p\!+\!1$ independent field components consisting of the $X^{\rm A}$ and of
$\varphi$ by the imposition of the Lagrangian displacement conditions (6.2) and
(6.3), and the further elimination by (6.4) of the $n\!-\!p\!-\!1$ degrees of freedom of
adjustment of the location of the $(p\!+\!1\!)$ dimensional world sheet in directions
orthogonal to itself is compensated by the introduction of the $n$ new degrees
of freedom of adjustment of the components of the relative displacement vector
$\DD^\mu$, the components $\zeta^\mu$ representing the  $p$ degrees of freedom
of the $X^{\rm A}$ together with the single degree of freedom of $\varphi$,
while the comonents $\xi^\mu$ correspond to the remaining $n\!-\!p\!-\!1$
degrees of freedom of adjustment of the brane locus. This means that for the
purposes of the action principle the independent infinitesimal variations of the
mappings (5.1), (5.2), and (5.3) are effectively and conveniently replaced, in a
Lagrangian reference system as characterised by (6.2), (6.3), (6.4), by  the
independent variation of the components just of the single infinitesimal
displacement vector $\DD^\mu$ or equivalently, in a semi Eulerian system as
characterised by (7.7), by the variation, subject the constraints (7.4) and
(7.12), of the separate vector fields $\zeta^\mu$ and $\xi^\mu$.

\vfill\eject

\bigskip\parindent=0 cm
{\bf 8.  The action for a ``warm" superfluid $p$-brane model.}
\medskip\parindent=2 cm

The action that we shall use has the form of a $p$-brane
world sheet integral 
$${\cal I}=\int {\cal L}\ d^{p+\!1}\!\sigma\ , \eqno(8.1)$$
in which the action density ${\cal L}$ has the form 
$${\cal L}=\Vert\eth\Vert^{_1\over^2}L \eqno(8.2)$$ 
where $\vert\eth\vert$ is the determinant of the internal $(p\!+\!1)$ surface metric
$\eth_{ij}$ as given by (5.7) and $L$ itself is a Lagrangian scalar of the form
(4.1). 

In evaluating the variation of this integral it will be most
instrictive to use the intermediate semi-Eulerian reference system,
in terms of which we shall have
$$\DS{\cal I}=\int (\DS{\cal L})\, d^{p+\!1}\!\sigma
 \eqno(8.3)$$
with
$$\DS{\cal L} =\Vert\eth\Vert^{_1\over^2}\diaS L \eqno(8.4)$$
where $\diaS$ is a modified pseudo differential operator given by
$$\diaS=\DS+{_1\over^2}\big(\eta^{\mu\nu}\DS g_{\mu\nu})\eqno(8.5)$$
whose introduction in place of $\DS$ allows us to work with exclusively
tensorial quantities by including an adjustment term allowing for the
variation (7.8) of the $(p\!+\!1)$ surface measure density factor  $\Vert
\eth\Vert^{_1\over^2}$.

For the evaluation of the variation in a compact region of the integral as a
whole it would of course make no difference whether we use a fully Lagrangian or
a semi-Eulerian reference system, i.e. we have
$$\DL{\cal I}=\DS{\cal I} \eqno(8.6)$$
since the difference in the corresponding integrands will just have the form of
a pure $(p\!+\!1)$ surface divergence, i.e.
$$\diaL L=\diaS  L+\ov\nabl_\nu\big(\zeta^\nu L)\eqno(8.7)$$
where $\zeta^\nu$ is the tangential displacement vector introduced
in the previous section.

Since $L$ is postulated to be just a covariant scalar function of the entropy
current vector $s^\nu$ as constructed in section 5, and of the momentum-energy
covector $\mu_\nu$ that is constructed in accordance with (2.1) as the gradient
of the phase scalar $\varphi$, it follows that its semi Eulerian differential
will be expressible in terms of a thermal momentum-energy covector $\Theta_\nu$
that is the dual of $s^\mu$ and of a condensate current vector $\nnu^\mu$ that
is the dual of $\mu_\nu$in the form
$$\DS L=\Theta_\nu\DS s^\nu-\nnu^\nu\DS\mu_\nu+{\mit 1\over 2}
\eta^{\mu\rho}\big(\Theta_\rho s^\nu+\mu_\rho \nnu^\nu)\DS g_{\mu\rho}
\eqno(8.8)$$
where the form of the coefficient of the metric variation in the final
term is easily derivable (as a Noether identity$^{[31]}$) from the
covariance condition, and where, in view of the tangentiality constraints
$$ s^\rho\omic\!^{\!\mu}_{\ \rho} =0\ ,
\hskip 2cm   \mu^\nu\omic\!^{\!\nu}_{\ \rho}=0\eqno(8.9)$$
to which $s^\rho$ and $\mu_\rho$ are subject by construction,
it is to be understood that the specification of the partial derivatives
is resolved by the imposition of corresponding tangentiality conditions
$$\Theta^\nu\omic\!^{\!\nu}_{\ \rho}=0\ , \hskip 2 cm
\nu^\rho\omic\!^{\!\mu}_{\ \rho} =0\ . \eqno(8.10)$$

It is to be remarked that just as the construction  of the current $s^\rho$
automatically ensures that it will satisfy the conservation law (5.8), so
analogously the construction of the energy-momentum covector $\mu_\rho$ as a
tangential gradient by (2.1) automatically ensures that it will satisfy a
corresponding conservation law as a consequence of the commutator identity
$$\ov\nabl_{[\mu}\ov\nabl_{\nu]}\varphi=
K_{[\mu}{^\rho}{_{\nu]}}\ov\nabl_\rho\varphi\eqno(8.11)$$ which must be
satisfied by any scalar field $\varphi$, where $K_\mu{^\rho}{_\nu}$ is the {\it
second fundamental tensor} (which determines both the inner and the outer
curvature$^{[21][36][37]}$ of the world sheet)  as defined$^{[15][18]}$ in terms
of the first fundamental tensor $\eta_\mu^{\ \nu}$ by the construction
$$K_{\mu\nu}{^\rho}=\eta_\mu^{\ \sigma}\ov\nabl_\nu\eta_\sigma^{\ \rho}
\eqno(8.12)$$
which not only ensures the obvious tangentiality and orthogonamity properties
$$\omic\!^{\!\mu}_{\ \sigma} K_{\sigma\nu}{^\rho} =0=
K_{\mu\nu}{^\sigma}\eta_\sigma^{\ \rho} \eqno(8.13)$$
but also (as a surface integrability condition) the non-trivial Weingarten
identity
$$ K_{[\mu\nu]}{^\rho}=0 \  .\eqno(8.14)$$
One  thus obtains a corresponding identity of the form
$$\ov\nabl_{[\mu}\mu_{\nu]}=K_{[\mu}{^\rho}{_{\nu]}}\mu_\rho
\eqno(8.15)$$
as the integrability condition for $\mu_\rho$ to have the form (2.1).

\vfill\eject

\bigskip\parindent=0 cm
{\bf 9.  Dynamics of a ``warm" superfluid $p$-brane model.}
\medskip\parindent=2 cm

To deal with the internal part of the variation what 
now has to be done
is to apply the translation relation (7.14) to the Lagrangian
formulae (6.8) and (6.9) for $s^\rho$ and $\mu_\rho$ we
obtain the corresponding semi Eulerian variations needed for
working out (8.8) in the form
$$\DS s^\rho=-{_1\over^2}s^\rho \eta^{\mu\nu}\DS g_{\mu\nu} +
s^\rho\ov\nabl_\nu\zeta^\nu-\zeta^\nu\ov\nabl_\nu s^\rho+
s^\nu\ov\nabl_\nu \zeta^\rho \eqno(9.1)$$
and
$$\DS \mu_\rho=\mu^\nu\omic\!^{\!\mu}_{\ \rho}\,
( \DS g_{\mu\nu}+\ov\nabl_\nu\zeta_\mu)
-\zeta^\nu\ov\nabl_\nu \mu_\rho
-\mu_\nu\ov\nabl_\rho \zeta^\nu \ . \eqno(9.2)$$
from which one immediately obtains
in variational integrand in  (8.4) is the
in the form
$$\diaS L={_1\over ^2}\ov T^{\mu\nu}\DS g_{\mu\nu}+\Theta_\rho\
\big(s^\nu\ov\nabl_\nu\zeta^\rho-\zeta^\nu\ov\nabl_\nu s^\rho\big)
+\nu^\rho\big(\mu_\nu\ov\nabl_\rho\zeta^\nu+\zeta^\nu\ov\nabl_\nu\mu_\rho\big)
 \eqno(9.3)$$
where the the coefficient $\ov T^{\mu\nu}$ of the metric variation,
which is of course to be interpreted as the surface {\it stress momentum
energy tensor}, is given by
$$\ov T^{\mu}_{\ \nu}=s^\mu\Theta_\nu +\nu^\mu \mu_\nu+\Psi\eta^\mu_{\ \nu}
\eqno (9.4)$$
in which the scalar potential $\Psi$, which is interpretable as a generalised
pressure function, has the form 
$$\Psi=L-s^\nu\Theta_\nu  \ .   \eqno(9.5)$$

To deal with the external part of the variation, i.e. the part concerned
with the displacement of the world sheet itself, it remains
to apply the translation relation (7.13) to the metric, which gives
$$\DS g_{\mu\nu}= \DE g_{\mu\nu}+2\nabla_{(\mu}\xi_{\nu)}.\eqno(9.6)$$
In order to proceed with the application of the variation principle without
having to worry about ultraviolet divergence and renormalisation effects, we now
make the postulate that the background spacetime remains fixed. This means
discounting  backreaction effects  due to emmission of gravitational radiation,
which will be an extremely good approximation not only for ``lightweight" string
models characterised by $Gm^2\approx 10^{-32}$ but even the ``heavyweight"
G.U.T. string models characterised by $Gm^2\approx 10^{-6}$ which is still
sufficiently small for it to be entirely satisfactory to ignore gravitational
radiation in the treatment of the short run dynamical behaviour, and to take
account of the backreaction phenomenon only as a mechanism of slow ``secular"
energy loss$^{[33][34][35]}$ that is of physical significance only as a statistically
accumulative effet in the very long run. This understanding allows us to take
the Eulerian part of the metric perturbation here simply to be zero,
$$ \DE g_{\mu\nu}=0 \eqno(9.7)$$
thereby completing the reduction of the problem to the required form
in which the only independent variations are those of the tangential
and orthogonal displacement vectors $\zeta^\nu$ and $\xi^\nu$.

We now come to the final step in the application of the variation principle
which is traditionally performed by combining the derivatives of the independent
variables into divergences that can be integrated out using Green's theorem
leaving only undifferentiated terms whose coefficients are the Eulerian
derivatives that must be set to zero as the dynamical equation. As far as the
derivatives tangential displacement variable $\zeta$ is concerned we shall now
go ahead in this traditional way using the symmetry condition
$$\nu^\mu\zeta^\nu\ov\nabl_{[\mu}\mu_{\nu]}=0 \eqno(9.8)$$
that is obtained from (8.15) using (8.13), but to deal with the derivatives of
the orthogonal displacement variable $\xi^\nu$ we shall use a trick of a less
traditional but even more effective kind, differentiating the orthogonality
condition  (7.12) itself, which gives the relation
$$\eta^\nu_{\ \rho}\ov\nabl_\mu \xi^\rho=-K^{\ \nu}_{\mu \ \rho}\xi^\rho
\eqno(9.9)$$
which allows {\it local} elimination (without any integration or use of Green's
theorem) of the derivatives of $\xi^\mu$. This strategy converts the variational
integrand in (8.4) to the final form
$$\diaS L=\ov T^{\mu\nu}K_{\mu\nu\rho}\xi^\rho -\zeta^\rho
\Big(2s^\mu\ov\nabl_{[\mu}\Theta_{\rho]}+\mu_\rho\ov\nabl_\mu\nu^\mu\Big)
+\nabl_\mu\big\{ 2\zeta^{[\rho}s^{\mu]}\Theta_\rho+\zeta^\rho\mu_\rho\nu^\mu
\big\} \ . \eqno(9.12)$$

It is now immediately apparent that the requirement that the the
corresponding integral variation (8.4) should vanish for an arbitrary
orthogonal displacement $\xi^\rho$ of the world sheet entails the external
force balance condition
$$\ov T^{\mu\nu}K_{\mu\nu\rho}=0   \ ,\eqno(9.13)$$
which is of a form that is applicable$^{[18][36]}$ to  any isolated
unpolarized brane model, the specific form of the stress momentum
energy tensor (as given by (9.4) in this particular case)
being all that distinguishes one kind of model from another.

Since the last term in (9.6) has the form of a worldsheet divergence
that integrates out,  the requirement that the integral variation
(8.4) should also vanish for an arbitrary tangential displacement
$\zeta^\nu$ evidently entails that the requisite internal dynamical equations
be given by
$$\eta_\nu{^\rho} \Big(2s^\mu\ov\nabl_{[\mu}\Theta_{\rho]}+
\mu_\rho\ov\nabl_\mu\nu^\mu\Big) =0\ . \eqno(9.14)$$
By contracting this with $s^\nu$ it can be seen (subject to the assumption 
implicit throughout his work that $s^\rho$ a nd $\mu_\rho$ are 
both strictly timelike so that their product
$s^\nu\mu_\nu$ will not vanish) that we are left with a condensate
surface current conservation law of the form 
$$\ov\nabl_\mu\nu^\mu=0 \ ,  \eqno(9.15)$$
which implies that the remaining content of (8.11) will reduce
to the form
$$\eta_\nu^{\ \rho}s^\mu\ov\nabl_{[\mu}\Theta_{\rho]}=0\ , \eqno(9.16)$$
which is the natural generalisation of the thermal momentum transport in the
ordinary superfluid case$^{[38]}$ to which it reduces in the case $p+1=n$ for
which $\eta_\nu^{\ \rho}$ is just the identity projector $g_\nu^{\ \rho}$.

\bigskip\parindent=0 cm
{\bf 10. Duality in the dynamics of warm string models.}
\medskip\parindent=2 cm

The outcome of the preceeding work was that he complete set of dynamical
equations of motion of the warm $p$ brane model would be given by the external
dynamical equation of motion (9.10), the internal dynamical equations (9.12) and
(9.13), together with the defining relation (2.1) for $\mu_\rho$ in terms of
$\varphi$, or equivalently the corresponding integrability condition (8.15)
whose essential surface projected part is just the symmetry condition
$$\eta_\mu^{\ \rho}\eta_\nu^{\ \sigma}\ov\nabl_{[\rho}\mu_{\sigma]}=0
\eqno(10.1)$$
and finally the entropy conservation law (5.12) which is formally
analogous to the dynamic current conservation law (9.12), but whose
status in the present variational formulation is merely that of a kinematic
identity.

Although the currents $s^\rho$ and $\nu^\rho$ have qualitatively very
different roles in the variational formulation used here,  which is the one
that arised most naturally from the context in which the concept of such a
``warm" brane model originated -- it is nevertheless to be emphasized that
there will be no such qualitative distinction within the physical ``on
shell" mechanical model that is ultimately obtained. It is manifestly
apparent$^{[39]}$ in the present relativistic version (and, though it was not
so obvious in the notation that was traditionally used in the past, can also
be demonstrated$^{[40]}$ in the original Newtonian version) of the Landau
superfluid model  that there is a strong analogy between
the roles of the entropy current $s^\rho$ and its dynamical conjugate, the
thermal momentum energy covector $\Theta_\rho$ on one hand, and the roles of
the condensate current $\nu^\rho$  and its dynamical conjugate, the
condensate momentum energy covector $\mu_\rho$ on the other hand. 

In the particular case of a string, with $p=1$,  to which we shall from this
point on again restrict our attention, the analogy that was discernible in the
ordinary fluid case becomes perfect, since the fact that the string world
sheet is only two dimensional means that the basic thermal dynamical
equation (9.16) will in this case reduce simply to an irrotationality
condition of the {\it same} form as the integrability condition (10.1) obeyed by
$\mu_\rho$, i.e. we shall be left simply with
$$\eta_\mu^{\ \rho}\eta_\nu^{\ \sigma}\ov\nabl_{[\rho}\Theta_{\sigma]}=0
\ . \eqno(10.2)$$
An immediate consequence of this is that just as $\mu_\rho$ is derivable
globally from the condensate phase scalar $\varphi$, so
also the thermal momentum covector will be at least locally derivable
from a formally analogous thermal gauge potential, $\vartheta$ say,
via an analogous relation of the form
$$\Theta_\rho=\ov\nabl_\rho\vartheta \ .\eqno(10.3)$$

In addition to this perfect ``chemical symmetry" between the roles of the two
independent constituents, the ``warm" string case is characterised by a ``dual
symmetry" of the kind whose existence was previously noticed$^{[23]}$ in the
``cold" case, but which is more interesting in the present case becase the
discrete chemical and dual symmetries combine to generate a 
forefold group of permutations among
two pairs of 4 algebraicly (though not dynamically)
independent conserved currents. The currents involved are the ordinary entropy
current $s^\rho$ and the condensate current $\nu^\rho$ that have already been
introduced, together with their respective duals, which are a thermal momentum
vector  $\star\!\Theta^\rho$ and a condensate momentum vector $\star\nu^\rho$
defined in terms of the background pullback
$${\cal E}^{\mu\nu}={\cal E}^{ij}x^\mu_{\ ,i}x^\nu_{\ ,j} \eqno(10.4)$$
of the internal worldsheet alternating tensor (as used in (5.7)) by the 
relations
$$\star\!\Theta^\rho={\cal E}^{\rho\nu}\Theta_\nu\ , \hskip 2 cm 
  \star\!\mu^\rho={\cal E}^{\rho\nu}\mu_\nu\ ,  \eqno(10.5)$$
which are of course invertible to give
$$\Theta_\rho={\cal E}_{\rho\nu} \star\!\Theta^\nu\ , \hskip 2 cm 
  \mu_\rho={\cal E}_{\rho\nu}\star\!\mu^\nu\ .  \eqno(10.6)$$

In terms of these dual variables the ``closure" conditions (10.1)
and (10.3) translate into ordinary surface current conservation laws,
so that the complete set of internal equations of motion for
the warm string model reduces to a set of 4 such conservation laws,
consisting of two mtually dual pairs i.e.
a thermal pair
$$\ov\nabl_\nu{\star\!\Theta}^\nu =0= \ov\nabl_\nu s^\nu \eqno(10.7)$$
and its chemical analogue, the condensate pair
$$\ov\nabl_\nu{\star\!\mu}^\nu =0=\ov\nabl_\nu \nu^\nu \ . \eqno(10.8)$$
The surface stress momentum energy tensor appearing in the 
equation (9.13) for the external motion will be expressible
in a form that is manifestly both chemically and dually symmetric as
$$\ov T^{\rho}_{\ \sigma}={\Lambda\over\Lambda-\Psi}\big(
s^\rho\Theta_\sigma+\nu^\rho\mu_\sigma\big) +
{\Psi\over\Psi-\Lambda}\big(\star\!\Theta^\rho\star\!s_\sigma
+\star\!\mu^\rho \star\!\nu_\sigma\big) \ , \eqno(10.9)$$
where the entropy and condensate current duals are given by
$$\star s_\rho={\cal E}_{\rho\sigma}s^\sigma\ , \hskip 2 cm 
  \star \nu_\rho={\cal E}_{\rho\sigma}\nu^\sigma\  \eqno(10.10)$$
while the dynamical conjugate $\Lambda$ of the pressure function
$\Psi$, is given by
$$\Lambda=L+n^\rho\mu_\rho\ ,\eqno(10.11)$$
so that by (9.5) we have
$$\Psi-\Lambda=-s^\rho\Theta_\rho -n^\rho\mu_\rho
=\star\!\Theta^\rho\star\!s_\rho
+\star\!\mu^\rho \star\!\nu_\rho\ . \eqno(10.12)$$
It is to be remarked that the (negative) scalar $\Lambda$ introduced in this way
could be used as a Lagrangian for an alternative variational formulation of
purely convective type$^{[37]}$ in which the independent currents $s^\nu$ and
$\nu^\nu$ are treated on the same footing.

\vfill\eject
\bigskip\parindent=0 cm
{\bf 11. Standard form of the stress momentum energy tensor.}
\medskip\parindent=2 cm

It is instructive to express the resutlts in terms of the standard
the preferred rest frame specified by the timelike
unit eigenvector $u^\rho$ of $\ov T^\nu_{\ \rho} $  which will have the
form $$u^\rho=(\Theta s+\mu \nu)^{-1}\big(\Theta\, s^\rho+\mu\, \nu^\rho\big)
\ ,\hskip 2 cm   u^\rho u_\rho=-1 \ ,\eqno(11.1)$$
and of the corresplonding spacelike unit eigenvector
$$v^\rho={\cal E}^{\rho\sigma}u_\sigma \ , \hskip 2 cm v^\rho v_\rho=1 
\ ,\eqno(11.2)$$
with the rest frame temperature $\Theta$ and the rest frame chemical
potential $\mu$ defined by
$$\Theta=-u^\rho\Theta_\rho\ , \hskip 2 cm \mu=-u^\rho\mu_\rho
\ , \eqno(11.3)$$
while the corresponding  rest frame number densities are defined by
 $$s=-u_\rho s^\rho\ , \hskip 2 cm  \nu=-u_\rho nu^\rho \ . \eqno(11.4)$$
In terms of this preferred fram
we can express the surface stress energy momentum tensor in the standard
form
$$ \ov T^{\rho\sigma}=U u^\rho u^\sigma-Tv^\rho v^\sigma \ , \eqno(11.5)$$
in which the eigenvalues $U$,  the mass-energy per unit length, and $T$, 
and the string tension, are obtainable from the relations
$$U+T=-\Lambda-\Psi \ ,\hskip 2 cm U-T=\Xi  \eqno(11.6)$$
where $\Xi$ is a positive scalar quantity $\Xi$ defined by
$$\Xi^2=\big(s^\rho\Theta_\rho-n^\rho\mu_\rho\big)^2+4s^\rho\mu_\rho
\nu^\sigma\Theta_\sigma \ ,\eqno(11.7)$$
from which it can be seen that the energy per unit length will be 
expressible as
$$ U=\Theta s+\mu \nu -\Psi \eqno(11.8)$$
while the corresponding dual formula giving the string tension will be
$$T=-\Lambda-\Theta s-\mu \nu \eqno(11.9)$$
(It is to be noted that in order for  $T$ to be positive like $U$ it is 
necessary for $\Lambda$ to be negative.)

\bigskip\parindent=0 cm
{\bf 12. Conserved numbers and equilibrium conditions.}
 \medskip \parindent=2cm

It follows from the the original current conservation laws (5.12) and
(9.15) for $s^\rho$ and $\nu^\rho$ that their dual  1-forms as defined by
(10.10) should satisfy irrotationality conditions analogous to
(10.1) and (10.2) and should therefore be locally derivable from
stream functions $S$ and $\psi$ say  in the form
$$ \star s_\rho=\ov\nabl_\rho S ,\hskip 2 cm  \star\nu_\rho=\nabl_
\rho\psi
\ ,\eqno(12.1)$$
where $S$ will evidently be a measure of the entropy as originally
introduced in section 5, while $\psi$ will be analogues measure
of the conserved flux associated with the condensate.
In the case of a closed string loop, these will determine corresponding
global circuit integrals
$$[S]=\oint \star\! s_\rho dx^\rho =\oint S_{,i}d\sigma^i \ ,
\hskip 2cm [\psi]=\oint \star\!\nu_\rho dx^\rho =\oint \psi_{,i}d\sigma^i \ ,
\eqno(12.2)$$
that will be conserved in the strong sense of being unaffected by
arbitrary continuous displacements of the circuit. Dually corresponding
globally conserved quantities are of course also obtained from the original
condensate phase scalar $\varphi$ as introdiced in (2.1) and its
thermal analogue as introduced in (10.3), in the form
$$[\varphi]=\oint \mu_\rho dx^\rho =\oint \varphi_{,i}d\sigma^i \ ,
\hskip 2cm [\psi]=\oint \star\Theta_\rho dx^\rho =\oint\vartheta_{,i}d\sigma^i
 \ , \eqno(12.3)$$
where $[\varphi]$ is interpretable as the total condensate phase winding
angle while $[\vartheta]$ is analogously interpretable as a thermal winding
angle. Thus whereas the loops of the simple ``hot" model described in section 3
and of the opposite extreme case of the ``cold" models described in section 2
are characterised by just two independent globally conserved quantities, namely
$[S]$ and $[\vartheta]$ in the ``hot" case, and $[\psi]$ and $[\varphi]$ in the
cold case, in the case of the more elaborate ``warm" models set up here a string
loop is characterised by all four of these quantities independently. 

It is the existence of such conserved quantities that makes it evident that (in
contrast with the degenerate special case of Goto Nambu string loops) these
various kinds of ``hot" ``warm" and ``cold" string loops cannot just disappear
by macroscopic radiation processes but, if left free from external perturbation,
will instead presumably tend to settle down in the towards stationary
equilibrium states  that minimise the mass energy ${\cal M}$ say for the given
values of the constants $[S]$, $[\vartheta]$, $[\psi]$, $[\varphi]$. Of course
in the very long run the assumptions on which these conservative ``hot" ``warm"
and ``cold" string models are based will cease to be exactly valid: entropy can
gradually be lost by microscopic thermal radiation processes, and although
topologically conserved in a classical description even the phase winding number
$[\varphi]$ in the ``cold" limit case can in principle decrease by quantum
tunnelling. The latter effect has already been investigated to some 
extent$^{[5]}$, what transpires being that the result is highly model dependent,
leading to decay timescales in some cases so short as to be comparable with the
dynamical timescales, in which case the classical string description breaks down
completely, but in other cases, still within physically realistic parameter
ranges, the tunnelling timescales are cosmologically long, leading, as noted in
the introduction, to the possibility of a catastrophic mass excess for strings
of the heavyweight type.  It is evident that allowance for the thermal effect
will tend to further enhance the long term survival capability of the string
loops, how much so being a subject for future work. 

The simplest configurations for the stationary ring states that will
ultimately be approached so long as the conservative description used
here remains valid will be of the circular rotating ring type that
first considered in detail by Davis and Shellard$^{[4][5][6]}$.
A wider study for generic string models  not only of circular equilibrium
states$^{[18][19]}$ but also of more general -- deformed -- 
equilibrium states$^{[20][21]}$
shows that the equilibrium of an isolated string loop 
in a flat background is characterised by a longitudinal running velocity
${\mit v}$ (which in the circular case is the product of the radius $r$ and the
angular velocity $\Omega$ say, i.e. ${\mit v}=r\Omega$) whose magnitude must be
the same as the extrinsic perturbation speed $c_{\rm E}$ as given
by the formula in (3.4) which remains valid in the generic case
(whereas the formula given for the longitudinal perturbation speed
$c_{\rm L}$ ceases to be valid for the ``warm" models
considered here, which, as in the ordinary
Landau superfluid, are characterised by not one but two distinct
``first" and ``second" longitudinal characteristic speeds). This means
that we can take over from our previous analysis the conclusion that
the longitudinal running speed in an equilibrium state will be given by
$$ {\mit v}^2= {T\over U} \ . \eqno(12.4)$$
As in the simpler cases considered previously the
intrinsic string state, which in this case depends on just
the three independents scalars that can be constructed from
$s^\rho$ and $\mu_\rho$, must be spacially as well as temporally 
{\it uniform},
since there will be no less than four independent Bernouilli type constants
obtained by contracting the Killing vector $k^\rho$ say generating the 
stationary symmetry with the four independent conserved momentum
1-forms, whose 
invariance with the stationary symmetry action is expressible, as a 
consequence of their internal irrotationality, by the uniformity conditions
$$\ov\nabl_\rho(k^\sigma\Theta_\sigma)=
\ov\nabl_\rho(k^\sigma\star\!s_\sigma)=\ov\nabl_\rho(k^\sigma\mu_\sigma)
=\ov\nabl_\rho(k^\sigma\star\!\nu_\sigma)=0 \ . \eqno(12.5)$$
Thus again as in the simpler cases considerered previously, the total
mass energy ${\cal M}$ will be expressible in terms of the
circumerence
$$\el=\oint d\el \ ,\hskip 2cm d \el^2=\eth_{ij}d \sigma^! d\sigma^j
\eqno(12.6)$$
(which in the circular case will be given by $\el=2\pi r$) and of the uniform
values of the energy per unit length $U$ (in the comoving rest frame) and of the
tension $T$ in the form 
$$ {\cal M} =(U+T)\ell \eqno(12.7)$$
while the magnitude ${\cal J}$ say of the angular momentum 
will be given by
$$\big({2\pi\cal J}\big)^2\leq UT \ell^4 \eqno(12.8)$$
with equality in the circular case for which the angular momentum takes the
maximum value allowed by the four global constants $[S]$ $[\vartheta]$ $[\psi]$
$[\varphi]$ characterising the loop, in terms of which the right hand side of
equation (12.8) will be given directly by
$$ U T\ell^4=\Big([\vartheta][S]+[\varphi][\psi]\Big)^2 \ .\eqno(12.9$$

\vfill\eject

\parindent=0 cm
{\bf Referencses}
\bigskip
 [1] N.K. Nielsen, P. Olesen, {\it Nucl. Phys.} {\bf B61} (1973) 45.
829 (1987). 
\smallskip

[2]  T.W.B. Kibble, {\it J.Phys.} {\bf A9} (1977) 1387.
\smallskip

[3] P. Peter, {\it Phys. Rev.} {\bf D46} (1992) 3322.
\smallskip

[4] R.L. Davis, E.P.S. Shellard, {\it Phys. Lett} {\bf B209} (1988) 485.
\smallskip

[5] R.L. Davis, {\it Phys. Rev} {\bf D38} (1988) 3722.
\smallskip

[6] R.L. Davis, E.P.S. Shellard, {\it Nucl. Phys.} {\bf B323} (1989) 209.
\smallskip

[7] B. Carter, in{\it Early Universe and Cosmic Structures 
(Xth Moriand Astrophysics Meeting)} ed A. Blanchard {\it et al}
 (Editions Fronti\`eres, Gif-sur-Yvette, 1991) 273.
\smallskip

[8] B. Carter, {\it Ann. N. Y. Acad. Sci.} {\bf 647} (1991) 758.
\smallskip

[9] J. Ostriker, C. Thompson, E. Witten {\it Phys.  Lett.} 
{\bf B180} (1986) 231.
\smallskip

[10] E. Copeland, M. Hindmarsh, N. Turok, {\it Phys. Rev. Lett.}
{\bf 58}, (1987) 1910.
\smallskip

[11] D. Haws, M. Hindmarsh, N. Turok, {\it Phys. Lett.} 
{\bf B209} (1988) 225.
\smallskip

[12] E. Copelans, D. Haws, M. Hindmarsh, N. Turok, {\it Nucl. Phys.}
{\bf B306} (1988) 908.
\smallskip

[13] C. Hill, H. Hodges, M. Turner, {\it Phys. Rev.} {\bf D37} (1988) 263.
\smallskip

[14]  A. Babul, T. Piran, D.N. Spergel, {\it Phys. Lett.}
{\bf B202} (1988) 207.
\smallskip

[15] B. Carter, {\it Phys. Lett.} {\it B228} (1989) 446.
\smallskip

[16]P. Peter, {\it Phys. Rev.} {\bf D46} (1992) 3335.
\smallskip

[17] P. Peter, {\it Phys. Rev.} {\bf D47} (1993) 3169.
\smallskip

[18] B. Carter, ``Covariant Mechanics of Simple ans Conducting Strings 
and Membranes'', in {\it The Formation and Evolution of Cosmic Strings},
ed G. Gibbons, S. Hawking, T. Vachaspati (Cambridge U.P., 1990)  213.
\smallskip

[19] B. Carter, {\it Phys. Lett.} {\it B238} (1990) 166.
\smallskip

[20] B. Carter, V.P. Frolov, O. Heinrich, {\it Class. and Quantum 
Grav.}  {\bf 8} (1991) 135.
\smallskip

[21] B. Carter, ``Basic Brane Mechanics",  in {\it Relativistic Astrophysics
and Gravitation  (proc 10th Potsdam Seminar, Oct. 1991)}
ed. S. Gottloeber, J.P. Muecket \& V. Mueller, 
 (World Scientific, Singapore, 1992) 300.
\smallskip

[22] E. Witten, {\it Nucl. Phys.} {\bf B249} (1985) 557.
\smallskip

[23] B. Carter, {\it Phys. Lett.} {\bf B224} (1989) 61.
\smallskip

[24] P. Fayet, {\it Nucl. Phys.} {\bf B347} (1990) 743.
\smallskip

[25] N. Dowrick, N.A. McDougal, {\it Phys. Rev.} {\bf D38} (1988) 3699.
\smallskip

[26] B. Carter, {\it Phys. Rev.} {\it D41} (1990) 3872.
\smallskip

[27] A. Vilenkin, {\it Phys. Rev.} {\it D41} (1990) 3038.
\smallskip

[28] N.K. Nielsen, {\it Nucl. Phys.} {\bf B167} (1980) 249.
\smallskip

[29] T.W.B. Kibble, ``Cosmic strings'', in 
{\it The Formation and Evolution of Cosmic Strings},
ed G. Gibbons, S.W. Hawking, T. Vachaspati (Cambridge U.P., 1990)  3.
\smallskip

[30] L.D. Landau, E.M. Lifshitz, {\it Course of Theoretical Physics,
Vol 6: Fluid Mechanics, Section 130} (Pergamon, Oxford,1959).
\smallskip

[31] B. Carter, I.M. Khalatnikov, {\it Phys. Rev. } {\bf D45} (1992) 4536.
\smallskip

[32] B. Carter, {\it Proc. Roy. Soc. Lond.} {\bf A433} (1991) 45.
\smallskip

[33] N. Turok, {\it Nucl. Phys.} {\bf B242} (1984) 520.
\smallskip

[34] T. Vachaspati, A. Vilenkin, {\it Phys. Rev.} {\it D31} (1985) 3035.
\smallskip

[35] R. Durer,  {\it Nucl. Phys.} {\bf B328} (1989) 238.
\smallskip

[36] B. Carter, {\it J. Geom. Phys.} {\bf 8} (1992) 53
\smallskip

[37] B. Carter, {\it Class. Qantum Grav.} {\bf 9} (1992) 19.
\smallskip

[38]  B. Carter, I.M. Khalatnikov, {\it Ann. Phys. } {\bf 219} (1992) 243.
\smallskip

[39]  B. Carter, I.M. Khalatnikov, {\it Rev. Math. Phys.} {\bf 6},
(1994) 277.
\smallskip

[40] B. Carter, ``Covariant Theory of Conductivity in Ideal Fluid or Solid Media",
in {\it Relativistic Fluid Dynamics},
ed.  A.M. Anile, \& Y. Choquet-Bruhat, Lecture Notes in Mathematics {\bf 1385}
(Springer - Verlag, Heidelberg, 1989) 1.

\end